\newlength{\extralineskip}
  \documentstyle[12pt]{article}
\def\stdi{\strut\displaystyle}
\def\ooo#1#2{{\stdi#1\over\stdi#2}}
\begin{document}
\begin{titlepage}
\begin{flushright}
          \begin{minipage}[t]{12em}
          \large UAB--FT--317\\
                 December 1993
          \end{minipage}
\end{flushright}

\vspace{\fill}

\vspace{\fill}

\begin{center}
\baselineskip=2.5em
{\huge CONSTRAINTS ON FERMION MAGNETIC AND\\
           ELECTRIC MOMENTS FROM LEP-I$^*$}\\
\end{center}

\vspace{\fill}

\begin{center}
{\sc R.ESCRIBANO$^{\dag}$ and E.MASS\'O}$^{\ddag}$\\
     Grup de F\'\i sica Te\`orica and Institut de F\'\i sica d'Altes Energies\\
     Universitat Aut\`onoma de Barcelona\\
     08193 Bellaterra, Barcelona, Spain
\end{center}

\vspace{\fill}

\begin{center}
\large  ABSTRACT
\end{center}
\begin{center}
\begin{minipage}[t]{36em}
	The effective Lagrangian approach allows us to
	constrain fermion magnetic and electric moments
	using LEP-I data. We improve some of the previous
	limits on these moments.
\end{minipage}
\end{center}

\vspace{\fill}

{\noindent\makebox[10cm]{\hrulefill}\\
\footnotesize
\makebox[1cm][r]{$^{\dag}$} Research supported by a F.P.I. grant from the
                            Universitat Aut\`onoma de Barcelona.
                                                                  \\
\makebox[1cm][r]{\        } E-mail address: {\tt escribano$@$ifae.es}\\
\makebox[1cm][r]{$^{\ddag}$} E-mail address: {\tt masso$@$ifae.es}\\[0.2cm]
\makebox[1cm][r]{     $^*$} Work partially supported by
the research project CICYT-AEN90-0019.
}
\end{titlepage}

\clearpage

%
\def\tableone{
\begin{table}
\centering
\begin{tabular}{|l|l|l|}\hline
$a_f$&
OUR LIMITS&
OTHER LIMITS   \\ \hline
\hline
$\tau$        & $0.0062$        & $0.02$              \cite{SILVERMAN}   \\
              &                 & $0.11$              \cite{GRIFOLS}     \\
              &                 & $0.39 \pm 0.30$     \cite{DOMOKOS}     \\
\hline
$\nu_\tau$    & $3.6\ 10^{-6}$  & $5.4\ 10^{-7}$      \cite{COOPER}      \\
              &                 & $4\ 10^{-6}$        \cite{CROTCH}      \\
              &                 & $8\ 10^{-6}$        \cite{GIUDICE}     \\
\hline
$u$           & $2.3\ 10^{-5}$  & $3\ 10^{-4}$        \cite{SILVERMAN}   \\
\hline
$d$           & $9.0\ 10^{-5}$  & $6\ 10^{-4}$        \cite{SILVERMAN}   \\
\hline
$c$           & $6.8\ 10^{-3}$  & $0.030$             \cite{SILVERMAN}   \\
\hline
$s$           & $1.8\ 10^{-3}$  & $0.025$             \cite{SILVERMAN}   \\
\hline
$b$           & $4.5\ 10^{-2}$  & $0.13$              \cite{SILVERMAN}   \\
              &                 & $-0.34 \pm 0.42$    \cite{DOMOKOS}     \\
\hline
\end{tabular}
\caption{}
\label{Table 1}
\vspace{2ex}
\end{table}
}
\def\tabletwo{
\begin{table}
\centering
\begin{tabular}{|l|l|l|}\hline
$d_f$&
OUR LIMITS&
OTHER LIMITS    \\
&(units {\it e-cm})&
(units {\it e-cm})   \\ \hline
\hline
$\tau$        & $3.4\ 10^{-17}$   & $1.2\ 10^{-16}$  \cite{BERNREUTHER}   \\
              &                   & $1.4\ 10^{-16}$  \cite{AGUILA,BARR}   \\
              &                   & $6\ 10^{-16}$    \cite{GRIFOLS}       \\
\hline
$\nu_\tau$    & $6.9\ 10^{-17}$   & $1.6\ 10^{-16}$  \cite{GIUDICE}       \\
\hline
$c$           & $8.9\ 10^{-17}$   & $6.5\ 10^{-23}$  \cite{DE RUJULA}     \\
\hline
$s$           & $8.9\ 10^{-17}$   & $1.0\ 10^{-24}$  \cite{DE RUJULA}     \\
\hline
$b$           & $8.9\ 10^{-17}$   & $7.8\ 10^{-21}$  \cite{DE RUJULA}     \\
\hline
\end{tabular}
\caption{}
\label{Table 2}
\end{table}
}
%
%
\def\figspageone{
\begin{figure}
\centering
\vspace{8cm}
\label{Figure 1 (a)}
\end{figure}
\vspace{2cm}
\begin{figure}
\centering
\vspace{8cm}
\label{Figure 1 (b)}
\end{figure}
}
\def\figspagetwo{
\begin{figure}
\centering
\vspace{8cm}
\label{Figure 1 (c)}
\end{figure}
\vspace{2cm}
\begin{figure}
\centering
\vspace{8cm}
\label{Figure 1 (d)}
\end{figure}
}
\def\figspagethree{
\begin{figure}
\centering
\vspace{8cm}
\label{Figure 1 (e)}
\end{figure}
\vspace{2cm}
\begin{figure}
\centering
\vspace{8cm}
\label{Figure 1 (f)}
\end{figure}
}
\def\figspagefour{
\begin{figure}
\centering
\label{Figure 2 (a)}
\end{figure}
\vspace{2cm}
\begin{figure}
\centering
\vspace{8cm}
\label{Figure 2 (b)}
\end{figure}
}
\def\figspagefive{
\begin{figure}
\centering
\vspace{8cm}
\label{Figure 2 (c)}
\end{figure}
\vspace{2cm}
\begin{figure}
\centering
\vspace{8cm}
\label{Figure 2 (d)}
\end{figure}
}
\def\figspagesix{
\begin{figure}
\centering
\vspace{8cm}
\label{Figure 2 (e)}
\end{figure}
\vspace{2cm}
\begin{figure}
\centering
\vspace{8cm}
\label{Figure 2 (f)}
\end{figure}
}

\addtolength{\baselineskip}{\extralineskip}
\vbox{\vspace{6em}}

\section{Introduction}

The magnetic and electric dipole moments of fermions contain important
information about their nature and interactions. On the one hand, concerning
the magnetic moments, the precise measurements of the $g-2$ values of the
electron and the muon provide, among other things, an accurate test of the
point-like character of both leptons. However, limits on the magnetic dipole
moment of the tau lepton, of the neutrinos and of the different quark species
are much poorer.

On the other hand, the most stringent limits on electric dipole moments
concern the neutron and the electron, and this sets strong constraints
on some of the CP-violating parameters. Using quark-model arguments,
one is able to infer upper bounds on the electric dipole moments of the up
and down quarks from the neutron measurements. However, again the electric
dipole moments of the tau lepton, of the neutrinos, and of the second and
third generation quarks are much less constrained.

The aim of the present paper is to extract ``indirect'' limits on fermion
moments using the electroweak data, in the context of an effective Lagrangian
approach. Our main idea is fairly simple. The effective Lagrangian that
may induce fermion moments different from the Standard Model expectations
ine\-vitably induces anomalous couplings of the neutral boson Z to fermions.
We will obtain constraints on these anomalous couplings using the available
electroweak data, and these contraints will provide bounds on the fermion
magnetic and electric moments.

Let us define here the general electromagnetic matrix element describing
the interaction of a fermion with the photon. For a charged fermion
one usually defines
\begin{equation}
   \begin{array}{rl}
      <p_2|\ J_{em}^\mu(0)\ |p_1>\ \ =&\!
           -e\, Q_f\, \bar u(p_2) \left( F_1^f\, \gamma^\mu\,
           + \ooo{i}{2 m_f}\, F_2^f\, \sigma^{\mu\nu}\,
           q_\nu\, \right) u(p_1) \\[3ex]
           +&\! e\, \bar u(p_2)\, F_3^f\, \gamma_5\,
           \sigma^{\mu\nu}\, q_\nu\, u(p_1) \ ,\\[1ex]
   \end{array}
\end{equation}
where $e\, Q_f$ is the fermion charge, and $q=p_2-p_1$. The form factors
$F_2$ and $F_3$, evaluated in the static limit, correspond to the
anomalous magnetic moment
\begin{equation}
      a_f = F_2^f (q^2 = 0) \ ,
\end{equation}
and to the electric dipole moment
\begin{equation}
      d_f =e\,  F_3^f (q^2 = 0) \ .
\end{equation}

The electromagnetic interactions of neutrinos are described in terms of
\begin{equation}
   <p_2|\ J_{em}^\mu(0)\ |p_1>\ =
      \, \bar u(p_2)\, \left( i\, F_2^\nu\, \mu_B +
      e\, F_3^\nu\, \gamma_5\, \right) \sigma^{\mu\nu}\,
      q_\nu\, u(p_1) \ ,
\end{equation}
with $\stdi{\mu_B=\ooo{e}{2m_e}}$ the Bohr magneton. In units of $\mu_B$,
the neutrino magnetic moment is
\begin{equation}
      \kappa_\nu = F_2^\nu (q^2 = 0) \ ,
\end{equation}
and the electric dipole moment is
\begin{equation}
      d_\nu = e\, F_3^\nu (q^2 = 0) \ .
\end{equation}

In the following section we discuss the general analysis of the fermion
moments in the linear effective Lagrangian approach, and in section 3
we describe our procedure to extract our limits and find them.
The non-linear effective Lagrangian approach is discussed in section 4.
In section 5 we present our main conclusions.


\section{Linear effective Lagrangian analysis of fermion moments}

Deviations from the electroweak Standard Model (SM) realized in its
minimal linear form, with a perturbative scalar sector, can
be treated by using effective Lagrangians. The general idea of
the linear effective Lagrangian approach is
that theories beyond the SM, emerging at some
characteristic energy scale $\Lambda$, have effects at
low energies $E \leq G_F^{-1/2}$, and these effects
can be taken into account by considering a Lagrangian
that extends the SM Lagrangian, $\cal L_{SM}$:
\begin{equation}
   \cal L = \cal L_{SM} + \cal L_{eff} \ .
\label{L}
\end{equation}
The effective Lagrangian $\cal L_{eff}$ contains operators of
increasing dimension that are built with the SM fields including
the scalar sector, and is
organized as an expansion in powers of $(1/\Lambda)$.

The success of the SM at the level of quantum corrections can
be considered as a check of the gauge symmetry properties of the
model. To preserve the consistency of the low energy theory, with
a Lagrangian given by Eq.\,(\ref{L}), we will assume that
$\cal L_{eff}$ is $SU(3) \otimes SU(2) \otimes U(1)$ gauge invariant.
Some of the problems that originate when dealing with non-gauge
invariant interactions have been discussed in \cite{EDU,KEK}.
The gauge-invariant operators that dominate at low
energies have dimension 6 and have been listed in \cite{BUCHMULLER & LEUNG}.
We will now write the set of operators that would
contribute to the dipole moments we are interested in.
For each one of the listed operators, its hermitian
conjugate will also contribute to the corresponding
moment. We will not display the hermitian conjugate,
although we considered it in our calculations.

Given a charged lepton $\ell$, there are two operators
contributing to its magnetic dipole moment
\begin{equation}
   \begin{array}{rl}
      \cal O_{\ell B}\ =&\! \bar L\, \sigma^{\mu\nu}\, \ell_R\,
                       \Phi\, B_{\mu\nu} \ ,\\[2ex]
      \cal O_{\ell W}\ =&\! \bar L\, \sigma^{\mu\nu}\,
                       \vec\sigma\, \ell_R\, \Phi\,
                       \vec W_{\mu\nu} \ ,\\[1ex]
   \end{array}
\end{equation}
where $L$ is the lepton isodoublet containing $\ell$ and
$\ell_R$ the singlet partner. As we will see, the operators
will involve the $U(1)$ and $SU(2)$ field strengths,
\begin{equation}
   \begin{array}{rl}
      B_{\mu\nu}\ =&\! \partial_{\mu} B_{\nu} - \partial_{\mu}
                   B_{\nu} \ ,\\[2ex]
      W_{\mu\nu}^i\ =&\! \partial_{\mu} W_{\nu}^i - \partial_{\nu}
                    W_{\mu}^i - g \epsilon^i_{jk}  W^j W^k \ ,\\[1ex]
   \end{array}
\end{equation}
as well as the Higgs field $\Phi$ or its conjugate $\tilde \Phi
=i \sigma_2 \Phi^*$.
When considering quarks we have to distinguish between up-type
and down-type induced magnetic moments. When the quark
is $U=u, c$ or $t$, we have contributions coming from
\begin{equation}
   \begin{array}{rl}
      \cal O_{UB}\ =&\! \bar Q\, \sigma^{\mu\nu}\, U_R\,
                   \tilde\Phi\, B_{\mu\nu} \ ,\\[2ex]
      \cal O_{UW}\ =&\! \bar Q\, \sigma^{\mu\nu}\,
                   \vec\sigma\, U_R\, \tilde\Phi\,
                   \vec W_{\mu\nu} \ ,\\[1ex]
   \end{array}
\end{equation}
where $Q$ is the corresponding quark isodoublet,
while for the case $D=d, s$ or $b$ one has
\begin{equation}
   \begin{array}{rl}
      \cal O_{DB}\ =&\! \bar Q\, \sigma^{\mu\nu}\, D_R\,
                   \Phi\, B_{\mu\nu} \ ,\\[2ex]
      \cal O_{DW}\ =&\! \bar Q\, \sigma^{\mu\nu}\,
                   \vec\sigma\, D_R\, \Phi\,
                   \vec W_{\mu\nu} \ .
   \end{array}
\end{equation}

Let us now display the operators contributing to the
electric dipole moments. For the charged leptons we
have
\begin{equation}
   \begin{array}{rl}
      \tilde{\cal O}_{\ell B}\ =&\! \bar L\, \sigma^{\mu\nu}\,
                               i\, \gamma_5\, \ell_R\, \Phi\,
                               B_{\mu\nu}\ ,\\[2ex]
      \tilde{\cal O}_{\ell W}\ =&\! \bar L\, \sigma^{\mu\nu}\,
                               i\, \gamma_5 \, \vec\sigma\, \ell_R\,
                               \Phi\, \vec W_{\mu\nu} \ ,\\[1ex]
   \end{array}
\end{equation}
while for $U=u, c, t$ we have
\begin{equation}
   \begin{array}{rl}
      \tilde{\cal O}_{UB}\ =&\! \bar Q\, \sigma^{\mu\nu}\,
                          i\, \gamma_5\, U_R\, \tilde\Phi\,
                          B_{\mu\nu}\ ,\\[2ex]
      \tilde{\cal O}_{UW}\ =&\! \bar Q\, \sigma^{\mu\nu}\, i\,
                          \gamma_5\, \vec\sigma\, U_R\, \tilde\Phi\,
                          \vec W_{\mu\nu} \ ,\\[1ex]
   \end{array}
\end{equation}
and finally, for $D=d, s, b$
\begin{equation}
   \begin{array}{rl}
      \tilde{\cal O}_{DB}\ =&\! \bar Q\, \sigma^{\mu\nu}\, i\,
                           \gamma_5\, D_R\, \Phi\, B_{\mu\nu} \ ,\\[2ex]
      \tilde{\cal O}_{DW}\ =&\! \bar Q\, \sigma^{\mu\nu}\,
                           i\, \gamma_5\, \vec\sigma\, D_R\, \Phi\,
                           \vec W_{\mu\nu} \ .
   \end{array}
\end{equation}

The case of the neutrino dipole moments has to be
treated separately. We need to enlarge the minimal
SM by adding the right-handed neutrinos $\nu_{eR},
\nu_{\mu R},\nu_{\tau R}$. Once this is done, we have
the following operators contributing to the neutrino
magnetic moment
\begin{equation}
   \begin{array}{rl}
      \cal O_{\nu_{\ell}B}\ =&\! \bar L\, \sigma^{\mu\nu}\,
                             \nu_{\ell R}\, \tilde\Phi\,
                             B_{\mu\nu} \ ,\\[2ex]
      \cal O_{\nu_{\ell}W}\ =&\! \bar L\, \sigma^{\mu\nu}\,
                             \vec\sigma\, \nu_{\ell R}\,
                             \tilde\Phi\, \vec W_{\mu\nu} \ ,\\[1ex]
   \end{array}
   \label{NU}
\end{equation}
and to the neutrino electric dipole moment
\begin{equation}
   \begin{array}{rl}
      \tilde{\cal O}_{\nu_{\ell}B}\ =&\! \bar L\, \sigma^{\mu\nu}\,
                                    i\, \gamma_5\, \nu_{\ell R}\,
                                    \tilde\Phi\, B_{\mu\nu} \ ,\\[2ex]
      \tilde{\cal O}_{\nu_{\ell}W}\ =&\! \bar L\, \sigma^{\mu\nu}\,
                                    i\, \gamma_5\, \vec\sigma\, \nu_{\ell R}\,
                                    \tilde\Phi\, \vec W_{\mu\nu} \ .\\[1ex]
   \end{array}
   \label{TILDENU}
\end{equation}
In (\ref{NU}) and (\ref{TILDENU}), $L$ is the lepton isodoublet
containing $\nu_\ell$.

The effective Lagrangian to be considered can now be written as a
linear combination of the operators we have listed
\begin{equation}
   \begin{array}{rl}
      \cal L_{eff}\ =&\! \displaystyle\sum_f \left({\alpha_{fB}
                    \over \Lambda^2}\ \cal O_{fB}
                    + {\alpha_{fW} \over \Lambda^2}\
                    \cal O_{fW} \right)\\[3ex]
                     +&\! \displaystyle\sum_f \left({\tilde\alpha_{fB}
                    \over \Lambda^2}\ \tilde{\cal O}_{fB}
                    + {\tilde\alpha_{fW} \over \Lambda^2}\
                    \tilde{\cal O}_{fW} \right) \ .\\[1ex]
   \end{array}
   \label{Leff}
\end{equation}

It is clear that below the scale of spontaneous symmetry
breaking the effective Lagrangian in Eq.\,(\ref{Leff})
induces contributions to the anomalous magnetic moments
$a_f$ and the electric moments $d_f$. Substituting
\begin{equation}
   \Phi \longrightarrow
        \left( \matrix{\stdi 0              \cr
               \stdi v \over \stdi \sqrt{2} \cr}
        \right) \ ,\\[1ex]
\end{equation}
with $v^2 = 1/(\sqrt{2} G_F) \simeq (246\ GeV)^2$, one
gets the following contributions
\begin{equation}
   \delta a_f\, =\, -\,2\, \sqrt{2}\ {m_f \over v}\ {1 \over
                {e\, Q_f}}\ \left( c_w\, \epsilon_{fB}
                + s_w\, \epsilon_{fW} \right) \ ,\\[1ex]
   \label{DELTA a_fU}
\end{equation}
for any charged fermion $f$ being the up-type component of the
lepton and quark isodoublets, and
\begin{equation}
   \delta a_f\, =\, -\,2\, \sqrt{2}\ {m_f \over v}\ {1 \over
                {e\, Q_f}}\ \left( c_w\, \epsilon_{fB}
                - s_w\, \epsilon_{fW} \right) \ ,\\[1ex]
   \label{DELTA a_fD}
\end{equation}
for the down-type components.
We have defined the dimensionless parameters
\begin{equation}
   \begin{array}{rl}
      \epsilon_{fB} = \alpha_{fB}\
                    {\stdi v^2 \over \stdi\Lambda^2} \ ,\\[1ex]
      \epsilon_{fW} = \alpha_{fW}\
                    {\stdi v^2 \over \stdi\Lambda^2} \ ,
   \end{array}
\end{equation}
valid for any fermion, and $c_w = cos\theta_w, s_w = sin
\theta_w$. When $f = \nu$ we obtain the following
contribution to the neutrino magnetic moment
\begin{equation}
   \delta \kappa_\nu\, =\, 2\, \sqrt{2}\ {m_e \over v}\ {1 \over
                       e} \left( c_w\, \epsilon_{\nu B}
                       + s_w\, \epsilon_{\nu W} \right) \ .
                       \\[1ex]
   \label{DELTA a_nu}
\end{equation}

Finally, for the up-type fermions $f$ one gets the following
contribution to $d_f$
\begin{equation}
   \delta d_f\, =\, -\,{\sqrt{2} \over v}\
                \left( c_w\, \tilde\epsilon_{fB}
                + s_w\, \tilde\epsilon_{fW} \right) \ ,\\[1ex]
   \label{DELTA d_fU}
\end{equation}
and for the down-type fermions
\begin{equation}
   \delta d_f\, =\, -\,{\sqrt{2} \over v}\
                \left( c_w\, \tilde\epsilon_{fB}
                - s_w\, \tilde\epsilon_{fW} \right) \ ,\\[1ex]
   \label{DELTA d_fD}
\end{equation}
where
\begin{equation}
   \begin{array}{rl}
      \tilde\epsilon_{fB} = \tilde \alpha_{fB}\
                             {\stdi v^2 \over \stdi\Lambda^2} \ ,\\[1ex]
      \tilde\epsilon_{fW} = \tilde \alpha_{fW}\
                             {\stdi v^2 \over \stdi\Lambda^2} \ .
   \end{array}
\end{equation}

We have discussed in this section how fermion moments are described
by linear effective Lagrangians. We will now address our attention to
describe the procedure to extract limits in this case, and leave the
discussion of non-linear effective Lagrangians until section 4.

\section{Bounds on the magnetic and electric moments}

We have seen in the last section that the linear effective Lagrangian in
Eq.\,(\ref{Leff}) induces fermion moments to be added to the SM contributions.
However, the new operators in the effective Lagrangian have other effects. The
most interesting, from the phenomenological point of view, is the anomalous
coupling of the Z boson to fermions
\begin{equation}
\begin{array}{rl}
\cal L\ =&\! \displaystyle\sum_{f=up-type} {\sqrt{2} \over v}\
             \left( s_w\, \epsilon_{f B} - c_w\, \epsilon_{f W} \right)\,
             \bar f\, \sigma^{\mu\nu}\, f\, \partial_\nu\, Z_\mu\\[1ex]
        +&\! \displaystyle\sum_{f=up-type} {\sqrt{2} \over v}\ \left( s_w\,
             \tilde\epsilon_{f B} - c_w\, \tilde\epsilon_{f W} \right)\,
             \bar f\, \sigma^{\mu\nu}\, i\, \gamma_5\, f\, \partial_\nu\,
             Z_\mu\\[1ex]
        +&\! \displaystyle\sum_{f=down-type} {\sqrt{2} \over v}\
             \left( s_w\, \epsilon_{f B} + c_w\, \epsilon_{f W} \right)\,
             \bar f\, \sigma^{\mu\nu}\, f\, \partial_\nu\, Z_\mu\\[1ex]
        +&\! \displaystyle\sum_{f=down-type} {\sqrt{2} \over v}\ \left( s_w\,
             \tilde\epsilon_{f B} + c_w\, \tilde\epsilon_{f W} \right)\,
             \bar f\, \sigma^{\mu\nu}\, i\, \gamma_5\, f\, \partial_\nu\,
             Z_\mu\  +\cdots \ .
\end{array}
\label{lefftot}
\end{equation}
These anomalous couplings would shift the partial widths
$\Gamma (Z \rightarrow f \bar f) \equiv \Gamma_f$ from
their predicted values in the Standard Model
\begin{equation}
      \Gamma_f = \Gamma_f^{SM} + \delta\Gamma_f \ .
\end{equation}

The agreement between experiment and the SM predictions
implies limitations on the strength of the different {\it
a priori} independent terms in the effective Lagrangian. We
will not allow for unnatural cancellations of the effects
produced by different operators, and thus we will consider
one operator at a time.

We should emphasize that once we have selected one of the
contributions to a particular fermionic width $\Gamma_f$,
it is not enough to simply use the experimental result
$\Gamma_f^{exp}$ to get our constraints. This is due to
the well-known $m_t$-dependence of the theo\-re\-tical
predictions of the SM, and to a less extent to the
dependence on the Higgs mass $M_H$. We choose to keep
the $m_t$-dependence of our results explicitily and allow
$M_H$ to span the range 60 GeV to 1 TeV, with this
``theoretical uncertainty'' (and the theoretical
uncertainty in the experimental value of $\alpha_s$)
linearly summed to the experimental errors. It is quite
common to combine theoretical and experimental errors
in quadrature. Here, we adopt the more conservative
point of view of adding the two types of uncertainties
{\em linearly}.

For the theoretical SM predictions we borrow the results of
Bardin {\it et al.\ }\cite{BARDIN}, that were kindly
made available to us by M.Bilenky.

We use the LEP value $M_Z = 91.187\ GeV$ as input, as well as
the current values of $\alpha$ and $G_F$
\cite{LEP,PDG}.
The observables we use to constrain deviations from the
Standard Model are:

\hspace{1.5cm} 1) the experimental LEP-I data
\cite{LEP}
\begin{equation}
   \begin{array}{rl}
      \Gamma_e      \ =&    83.86 \pm 0.30\ MeV    \ ,\\[1ex]
      \Gamma_\mu    \ =&    83.78 \pm 0.40\ MeV    \ ,\\[1ex]
      \Gamma_\tau   \ =&    83.50 \pm 0.45\ MeV    \ ,\\[1ex]
      \Gamma_{inv}  \ =&    497.6 \pm 4.3\ MeV     \ ,\\[1ex]
      \Gamma_{had}  \ =&    1740.3 \pm 5.9\ MeV    \ ,\\[1ex]
      g_{A_e}       \ =&    -0.50096 \pm 0.00093    \ ,\\[1ex]
      g_{A_\mu}     \ =&    -0.5013 \pm 0.0012    \ ,\\[1ex]
      g_{A_\tau}    \ =&    -0.5005 \pm 0.0014    \ ,\\[1ex]
   \end{array}
   \label{OBS}
\end{equation}

\hspace{1.5cm} 2) the $W$-mass determined from the ratio
$M_W/M_Z$ measured at $p \bar p$ colliders and the LEP value
for $M_Z$ \cite{ALLITI & ABE}
\begin{equation}
      M_W = \stdi 80.24 \pm 0.09^{+0.01}_{-0.02}\ GeV \ ,
            \\[1ex]
   \label{MW}
\end{equation}

\hspace{1.5cm} 3) the ratio of inclusive neutral- to
charged-currents neutrino cross sections on approximately
isoscalar targets \cite{ABRAMOWICZ & ALLABY}
\begin{equation}
   R_\nu = {\sigma \left(\nu N \rightarrow \nu
            +\cdots \right) \over
            \sigma \left(\nu N \rightarrow \mu
            +\cdots \right)} = 0.308 \pm 0.002 \ ,
   \label{RNU}
\end{equation}
and

\hspace{1.5cm} 4) the CDF limit on the top quark mass
\cite{CDF}
\begin{equation}
      m_t \geq 108\ GeV \ .
   \label{MTOP}
\end{equation}

We now turn our attention to the explicit
expressions for $\delta\Gamma_f$ due to the novel
effects we are considering. Starting with the
operators $\cal O_{fB}$, we have that for each
fermion $f$ the relative width shift is
\begin{equation}
      {\delta\Gamma_f \over \Gamma_f} =
      {s_w^2 \over {v_f^2 + a_f^2}}\ \epsilon_{fB}^2 \ ,
   \label{EPSfB}
\end{equation}
where $v_f = T_3^f - 2\, s_w^2\, Q_f$ and $ a_f = T_3^f$.
The operators $\cal O_{fW}$ induce the change
\begin{equation}
      {\delta\Gamma_f \over \Gamma_f} =
      {c_w^2 \over {v_f^2 + a_f^2}}\ \epsilon_{fW}^2 \ .
   \label{EPSfW}
\end{equation}

In the CP-violating sector, the operators
$\tilde{\cal O}_{fB}$ lead to
\begin{equation}
      {\delta\Gamma_f \over \Gamma_f} =
      {s_w^2 \over {v_f^2 + a_f^2}}\
      \tilde\epsilon_{fB}^2 \ ,
   \label{TILDE EPSfB}
\end{equation}
while the type $\tilde{\cal O}_{fW}$ lead to
\begin{equation}
      {\delta\Gamma_f \over \Gamma_f} =
      {c_w^2 \over {v_f^2 + a_f^2}}\
      \tilde\epsilon_{fW}^2 \ .
   \label{TILDE EPSfW}
\end{equation}

As we have already discussed, the limits on the
different parameters $\epsilon$ in Eqs.\,(\ref{EPSfB},\ref{EPSfW})
and $\tilde\epsilon$ in Eqs.\,(\ref{TILDE EPSfB},\ref{TILDE EPSfW})
will depend sensitively on $m_t$, since we have choosen to combine
the weak dependence on $M_H$ linearly with the
experimental uncertainties.
The limits are obtained by comparing
the experimental values of the observables
in Eqs.\,(\ref{OBS}--\ref{MTOP})
with the predictions that we get when adding the SM
results and the corrections previously discussed.
Our results on $\epsilon_{fB}$ are presented in Fig.\,1,
and the ones on $\epsilon_{fW}$ in Fig.\,2.
The projection on the $\epsilon^2$ (or the $m_t^2$)
axis corresponds to a single-variable $68{\%}$,
or $1\sigma$, confidence-level interval. These
limits can be read from Figs.\,1 and 2
\begin{equation}
    \begin{array}{ll}
        \epsilon_{eB}^2    \ \leq \ \ 1.10 \times 10^{-2} \hspace{5em}
       &\epsilon_{eW}^2    \ \leq \ \ 2.98 \times 10^{-3} \ ,\\[1ex]
        \epsilon_{\mu B}^2 \ \leq \ \ 1.13 \times 10^{-2} \hspace{5em}
       &\epsilon_{\mu W}^2 \ \leq \ \ 3.05 \times 10^{-3} \ ,\\[1ex]
        \epsilon_{\tau B}^2\ \leq \ \ 1.07 \times 10^{-2} \hspace{5em}
       &\epsilon_{\tau W}^2\ \leq \ \ 2.89 \times 10^{-3} \ ,\\[1ex]
        \epsilon_{UB}^2    \ \leq \ \ 7.17 \times 10^{-2} \hspace{5em}
       &\epsilon_{UW}^2    \ \leq \ \ 1.93 \times 10^{-2} \ ,\\[1ex]
        \epsilon_{DB}^2    \ \leq \ \ 7.17 \times 10^{-2} \hspace{5em}
       &\epsilon_{DW}^2    \ \leq \ \ 1.93 \times 10^{-2} \ ,\\[1ex]
        \epsilon_{\nu B}^2 \ \leq \ \ 4.31 \times 10^{-2} \hspace{5em}
       &\epsilon_{\nu W}^2 \ \leq \ \ 1.16 \times 10^{-2} \ .\\[1ex]
    \end{array}
    \label{EPSILON}
\end{equation}

As expected from a comparison of Eqs.\,(\ref{EPSfB}) and
(\ref{EPSfW}) the limits coming from the operators $O_{fW}$
are stronger than the ones coming from $O_{fB}$. Introducing the
limits on $\epsilon$ into Eqs.\,(\ref{DELTA a_fU},\ref{DELTA a_fD})
we get limits on the anomalous magnetic moments. Looking at this equation,
we anticipate that the operators $O_{fW}$ will lead to the
tightest bounds on $\delta a_f$. We only quote here these
strongest limits (we use $m_u=5\ MeV, m_d=10\ MeV, m_c=1.5\ GeV,
m_s=200\ MeV, m_b=5.0\ GeV$)
\begin{equation}
   \begin{array}{rl}
      \delta a_\tau             \ \leq& 6.2\times 10^{-3}   \ ,\\[1ex]
      \delta a_u                \ \leq& 2.3\times 10^{-5}   \ ,\\[1ex]
      \delta a_d                \ \leq& 9.0\times 10^{-5}   \ ,\\[1ex]
      \delta a_c                \ \leq& 6.8\times 10^{-3}   \ ,\\[1ex]
      \delta a_s                \ \leq& 1.8\times 10^{-3}   \ ,\\[1ex]
      \delta a_b                \ \leq& 4.5\times 10^{-2}   \ ,\\[1ex]
      \delta \kappa_{\nu_\tau}  \ \leq& 3.6\times 10^{-6}   \ .\\[1ex]
   \end{array}
\end{equation}
We have skipped the limits on $\delta a_e, \delta a_\mu$ since the
direct measurements make our limits completely useless. The
constraints on $\kappa_\nu$ for $\nu_e$ and $\nu_\mu$
coming from Red Giant observations \cite{RAFFELT} also make
our corresponding limits far away from being competitive.

We follow an identical procedure to constrain $\tilde\epsilon_{fB}$
and $\tilde\epsilon_{fW}$. It turns out that the limits on
$\tilde\epsilon_{fB}$ are the same than the ones on $\epsilon_{fB}$,
and similarly the $\tilde\epsilon_{fW}$ and $\epsilon_{fW}$ are
identical. Again we only quote our best limits on the electric
dipole moments (in {\it e-cm})
\begin{equation}
   \begin{array}{rl}
      \delta d_\tau           \ \leq& 3.4\times 10^{-17}     \ ,\\[1ex]
      \delta d_c              \ \leq& 8.9\times 10^{-17}     \ ,\\[1ex]
      \delta d_s              \ \leq& 8.9\times 10^{-17}     \ ,\\[1ex]
      \delta d_b              \ \leq& 8.9\times 10^{-17}     \ ,\\[1ex]
      \delta d_{\nu_\tau}     \ \leq& 6.9\times 10^{-17}     \ .\\[1ex]
   \end{array}
\end{equation}
Here we have skipped the limits on $\delta d_e, \delta d_\mu, \delta d_u$
and $\delta d_d$ since the experimental bounds on them are much
stronger. Again Red Giant evolution analysis imply much tighter
limits for $\nu_e$ and $\nu_\mu$ than our corresponding results.

We finally turn our attention to a comparison of our limits with
other results in the literature. We present the comparison in
Table \ref{Table 1} for the anomalous magnetic moment and in
Table \ref{Table 2} for the electric dipole moment.
We improve previous limits for
$a_\tau,a_u,a_d,a_c,a_s,a_b,d_\tau,d_{\nu_\tau}$.
Bounds on the tau-lepton moments following the approach presented
in this section were calculated in Ref.\,\cite{REM}.

\section{Non-linear effective Lagrangian approach}

In section 2 we discussed the case where we extend a linearly realized
Standard Model. However, there are models --as the technicolor ones--
that do not fit into the linear framework.
The assumptions of avoiding the presence of the physical scalar sector,
or of sending its mass to infinity do not commute with the linear
expansion since $\Phi$ is no longer there to construct gauge-invariant
effective operators.
In this case, $\Lambda \sim 4\pi v$, and it is appropriate
to use a non-linear or chiral realization of the Standard Model, and to
extend it using non-linear effective Lagrangians. These were discussed
in detail back in 1980 in \cite{CHIRAL} and have been recently reviewed
in \cite{FERUGLIO}.

We would like to investigate the consequences of adopting the chiral
approach for the fermion moments. Let us start with the magnetic moment
of the up and down quarks. By working this specific example we will
be able to understand the general pattern.

We add a piece to the SM Lagrangian that contains vertices corresponding
to $a_u$ and $a_d$. The leading terms in the chiral expansion are
given by
\begin{equation}
   \begin{array}{rl}
      \cal L_{eff}\ =&\! \displaystyle\ooo{g'\,\beta'}{v^2}\
		    \overline\psi_R\, M\, \hat{B}^{\mu\nu}\,
		    \sigma_{\mu\nu}\, \Sigma\dag\, \psi_L\\[2ex]
		    +&\! \displaystyle\ooo{g\,\beta}{v^2}\
                    \overline\psi_R\, M\, \Sigma\dag\, \hat{W}^{\mu\nu}\,
                    \sigma_{\mu\nu}\, \psi_L\ +\,h.c.\ ,\\[1ex]
   \end{array}
   \label{NlLeff}
\end{equation}
where
\begin{equation}
   \begin{array}{ccc}
	\psi_{L,R} &=& \left( \begin{array}{c}
				u \\ d
			      \end{array} \right)_{L,R}\ ,\\[3ex]
	\hat{W}_{\mu\nu} &=& \vec{W}_{\mu\nu}\cdot\vec\sigma\ ,\\[1ex]
	\hat{B}_{\mu\nu} &=& B_{\mu\nu}\cdot\sigma_3\ ,\\[1ex]
   \end{array}
\end{equation}
$M$ is the mass matrix
\begin{equation}
	\left( \begin{array}{cc}
		m_u & 0 \\
		0 & m_d
	       \end{array} \right)\ ,\\[1ex]
	\label{mass}
\end{equation}
and $\Sigma$ contains the would-be Goldstone bosons
\begin{equation}
	\Sigma = exp\,\{\,i\, \ooo{\vec\xi\cdot\vec\sigma}{v}\,\}\ .
\end{equation}

The contributions to the magnetic dipole moment coming from $\cal L_{eff}$
will contain the coefficients $\beta$ and $\beta'$ in Eq.\,(\ref{NlLeff}).
In the unitary gauge, these contributions can be put in the form of those
we found in the linear case,
Eqs.\,(\ref{DELTA a_fU},\ref{DELTA a_fD},\ref{DELTA a_nu},\ref{DELTA
d_fU},\ref{DELTA d_fD}),
provided we identify
\begin{equation}
	\begin{array}{ccc}
		\epsilon_{uB} &\rightarrow& \ooo{\sqrt{2}}{v}\
				g'\, \beta'\, m_u\ ,\\[1ex]
                \epsilon_{uW} &\rightarrow& \ooo{\sqrt{2}}{v}\
                                g\, \beta\, m_u\ ,\\[1ex]
                \epsilon_{dB} &\rightarrow& -\,\ooo{\sqrt{2}}{v}\
                                g'\, \beta'\, m_d\ ,\\[1ex]
                \epsilon_{dW} &\rightarrow& \ooo{\sqrt{2}}{v}\
                                g\, \beta\, m_d\ .\\[1ex]
	\end{array}
	\label{rdefeps}
\end{equation}

For our phenomenological purposes, it is interesting to notice that with
the very same identifications made in the last equation, our
Eq.\,(\ref{lefftot})
containing the anomalous couplings to the $Z$-boson is valid. We can use
now the fact, shown in the last section, that our constrains on the
anomalous magnetic moment from LEP-1 data do not depend on the definitions
of the parameters $\epsilon$. This is also true in the case of the dipole
electric moments, where we should use the effective Lagrangian
\begin{equation}
   \begin{array}{rl}
      \cal L_{eff}\ =&\! \displaystyle\ooo{g'\,\tilde\beta'}{v^2}\
                    \overline\psi_R\, M\, \hat{B}^{\mu\nu}\,
                    \sigma_{\mu\nu}\,i\, \gamma_5\,
		    \Sigma\dag\, \psi_L\\[2ex]
                    +&\! \displaystyle\ooo{g\,\tilde\beta}{v^2}\
                    \overline\psi_R\, M\, \Sigma\dag\, \hat{W}^{\mu\nu}\,
                    \sigma_{\mu\nu}\,i\, \gamma_5\,
		    \psi_L\ +\,h.c.\ .\\[1ex]
   \end{array}
   \label{NltLeff}
\end{equation}

Here we have worked out the example of the magnetic moments of the
$u$ and $d$ quarks. It is clear that it can be easily extended to the
magnetic and electric moments of all fermions. Notice also that our final
limits do not depend on the mass appearing in the mass matrix in
Eq.\,(\ref{mass}).

Our conclusion is that our limits on fermion magnetic and electric
moments derived in a linear effective Lagrangian approach hold true
when using the chiral expansion of the non-linear effective Lagrangian.

\section{Conclusions}

Physics beyond the Standard Model of the electroweak interactions can
be described in a general way using effective Lagrangians.
We use this approach to study magnetic and electric moments of the
fermions. We find that any contribution to these moments inevitably
induces anomalous couplings of the $Z$-boson to fermions.

The LEP-I data severely constrains these anomalous couplings, and
the constraints in turn set stringent bounds on the magnetic and
electric moments. Some of the bounds represent an improvement on
previously derived limits on fermion moments.

We have studied both the linear and non-linear realizations of
effective Lagrangians and have shown that both lead to the same
numerical limits on moments.

\section*{Acknowledgments}

One of us (R.E.) would like to thank M. Mart\'{\i}nez and F. Teubert
for their comments and helpful collaboration in providing
the most recent LEP-I data. A special thank goes to C. Ayala and
T. Coarasa for \LaTeX\ macros used in this work. I also
acknowledge financial support by a F.P.I. grant from the Universitat
Aut\`onoma de Barcelona.

%
%

\vspace{2em}

\section*{Table Captions}
\begin{description}

\item[Table 1:] Values of the anomalous magnetic moments excluded
	        by our analysis using LEP-I data, compared to other
		limits we have found in the literature.

\item[Table 2:] Same than Table 1 for the electric moments.

\end{description}
\vspace{2em}

\section*{Figure Captions}
\begin{description}

\item[Figure 1:] Allowed regions in the $(m_t^2,\epsilon^2_{fB})$
	         plane for $f=e$ {\bf (a)}, $f=\mu$ {\bf (b)},
		 $f=\tau$ {\bf (c)}, $f=U$ {\bf (d)},
		 $f=D$ {\bf (e)}, $f=\nu$ {\bf (f)}. Projection
		 on the $\epsilon^2$ axis corresponds to a $1\sigma$
		 CL interval.

\item[Figure 2:] Same than Figure 1 for the parameters $\epsilon^2_{fW}$.

\end{description}

  \clearpage

  \vspace{4em}

  \tableone

  \tabletwo

  \clearpage












\end{document}